\documentclass[twocolumn,showpacs,amsmath,amssymb,nofootinbib,superscriptaddress]{revtex4-1}
\usepackage{graphicx}
\usepackage{verbatim}
\newcommand{\ket}[1]{\ensuremath{|#1 \rangle}}
\newcommand{\ketbra}[2]{\ensuremath{| #1 \rangle\hspace{-2pt} \langle #2 |}}
\begin{document}
\title{Noisy quantum walks of two indistinguishable interacting particles}
\author{Ilaria Siloi}
\affiliation{Dipartimento di Scienze Fisiche, Informatiche e Matematiche, Universit\`a di Modena e Reggio Emilia, I-41125 Modena, Italy}
\author{Claudia Benedetti}
\affiliation{Quantum Technology Lab, Dipartimento di Fisica, Universit\`a  degli Studi di Milano, I-20133, Milano, Italy}
\author{Enrico Piccinini}
\affiliation{Dipartimento di Ingegneria dell'Energia Elettrica e
dell'Informazione ``Guglielmo Marconi" - DEI, Universit\`a di Bologna,
I-40135 - Bologna,Italy}
\author{Jyrki Piilo}
\affiliation{Turku Centre for Quantum Physics, Department of Physics and
Astronomy, University of Turku, FI-20014, Turun Yliopisto, Finland}
\author{Sabrina Maniscalco}
\affiliation{Turku Centre for Quantum Physics, Department of Physics and
Astronomy, University of Turku, FI-20014, Turun Yliopisto, Finland}
\author{Matteo G. A. Paris}
\affiliation{Quantum Technology Lab, Dipartimento di Fisica,
Universit\`a  degli Studi di Milano, I-20133, Milano, Italy}
\affiliation{INFN, Sezione di Milano, I-20133, Milano, Italy}
\affiliation{Centro S3, CNR-Istituto Nanoscienze,  I-41125 Modena, Italy}
\author{Paolo Bordone}
\affiliation{Dipartimento di Scienze Fisiche, Informatiche e
Matematiche, Universit\`a di Modena e Reggio Emilia, I-41125 Modena,
Italy} \affiliation{Centro S3, CNR-Istituto Nanoscienze,  I-41125
Modena, Italy}
\date{\today}
\begin{abstract}
We investigate the dynamics of continuous-time two-particle quantum
walks on a one-dimensional noisy lattice. Depending on the initial
condition, we show how the interplay between particle
indistinguishability and interaction determines distinct propagation
regimes. A realistic model for the environment is considered by
introducing non-Gaussian noise as time-dependent fluctuations of the
tunnelling amplitudes between adjacent sites. We observe that the
combined effect of particle interaction and fast noise (weak coupling
with the environment) provides a faster propagation compared to the
noiseless case. This effect can be understood in terms of the
band-structure of the Hubbard-model, and a detailed analysis as a
function of both noise and system parameters is presented.
\end{abstract}
\maketitle
\section{Introduction}
Quantum walks (QWs), the quantum counterpart of classical random walks,
describe the stochastic propagation of a quantum system (e.g. a
particle) on a discrete $n$-dimensional graph
\cite{kempe03,aharonov93,Va2012}.  A graph is any object that can be
mathematically described as a set of vertices (or sites) and edges (or
links between the sites).  The simplest graph to analyze the dynamics of
a QW is the one-dimensional lattice, i.e. the line
\cite{schreiber11,peter03,moqadam2015}, though more complex structures
have been also investigated \cite{kempe05,Xu08, Salimi09, volta09,
faccin13, adi14,mul14,difranco15,bisio16,mirc16} in order to fully characterize
the dynamics of QWs.  In particular, it has been shown that the final
state of the quantum walker  strongly depends on its initial conditions
and, because of  quantum interference, it generally propagates faster
than its classical counterpart. QWs have been extensively analyzed in
different contexts, ranging from quantum-enhanced search algorithms
\citep{childs04, optimal16} and universal models of quantum computation
\citep{childs09}, to quantum transport in complex networks
\citep{mulken11}, e.g. biological 
systems \citep{mohseni08,cho05, plenio08}.
\par
The study of few-particle QWs may offer a bottom-up
approach for understanding and simulating many body systems
\citep{lahini12,ahlbrecht12, crespi13}. In fact, besides photons
\cite{wang13,defienne16}, QWs have been implemented in many experimental
platforms such as trapped ions \citep{schmitz09,zahringer10} and neutral
atoms \citep{karski09,fukuhara13}; very recently a controlled dynamics
of two particle with tunable interaction has been demonstrated with
optical lattices \cite{preiss15}.
\par
As a matter of fact, the propagation of multiple 
indistinguishable particles is affected by the exchange symmetry 
even in the absence of interaction. This phenomenon, known as 
Hanbury Brown-Twiss (HBT) interference, may create non trivial
spatial correlations between two identical particles, and has been
widely investigated both experimentally \citep{bromberg09,peruzzo10} and
theoretically \citep{qin14,sansoni12,benedetti2012}. In turn the
evolution of free particles strongly depends on the statistics: while
bosons tend to propagate along the same direction an effect known as
{\em bunching}, fermions tend to move in the opposite directions, 
showing {\em antibunching}, and they have zero-probability to 
occupy the same site, consistently with the Pauli exclusion principle.  
Upon introducing interaction between particles, the picture becomes 
more involved. As predicted by the Bose-Hubbard Hamiltonian, 
stable repulsively bound pair has been observed \citep{winkler06}; 
moreover, under proper initial condition, the interplay between 
interaction and indistinguishability induces a continuous transition 
from bosonic- to fermionic-like spatial correlations \citep{lahini12}. 
All these effects have been shown to depend on the
strength of the interaction, but not on whether it is attractive or
repulsive, since the change in sign of the interaction $U$ simply
reverses the energy spectrum \citep{winkler06}. 
\par
In this framework, although there are some studies 
investigating the impact of decoherence and disorder on the 
dynamics of two-particle quantum walks
\citep{schreiber11,de14,lahini08,lahini10,lahini11,derrico13,beggi2016},
the combined effect of indistinguishability and interaction in a
classical noisy environment is still poorly understood.
In this paper, we aim to contribute to a better understanding
of the dynamics of this kind of systems.
In particular, we analyze in details the role 
of interaction in the propagation of two identical
particles hopping on a one-dimensional noisy lattice and 
discuss the interplay between interaction and 
indistinguishability in the presence of noise.
A realistic model for the QW environment is introduced,  
where the induced noise is described by non-Gaussian  
stochastic, time-dependent, fluctuations in the 
tunnelling amplitudes \citep{benedetti16,benedetti2013}. 
Upon tuning the spectral parameters of the noise, we explore 
different regimes ranging from the localization of the pair 
in the presence of {\it slow} noise, to non-ballistic propagation 
due to  {\it fast} noise. 
\par
Our results show that, in the ideal case of absence of noise, 
the strength of the interaction determines distinct propagation 
regimes. On the other hand, noise makes such distinction less sharp
and creates an intermediate regimes with a non trivial dynamics, 
which will be analyzed in details in our work.  We observe that
noise with a fast-decaying autocorrelation function induces a transition
from ballistic to diffusive propagation in the case of two
non-interacting walkers while, under proper initial conditions, noise
allows two interacting particles to propagate faster with respect to the
noiseless ballistic case.  We show that this phenomenon depends both on
the noise and the system parameters, 
and that it can be understood in terms of the band-structure of the
Hubbard model.
\par 
The paper is organized as follows: in Sec. \ref{sec2} we introduce 
our model for a two-particle continuous-time quantum walk
(CTQW) on a noisy lattice, whereas in Sec. \ref{sec3} we illustrate  
the dynamics of two interacting indistinguishable particles. 
We first show results for the noiseless case in Sec.  \ref{noiseless} 
and then, in Sec. \ref{noisy}, we illustrate the dynamical properties
of the system in the presence of noise. 
Section \ref{sec4} closes the paper with some concluding remarks.
%%%%%%%%%%%%%%%%%%%%%%%%%
\section{Model}
\label{sec2}
The continuous-time quantum walk \cite{farhi98, childs02} can be seen as the quantum version of
the classical continuous-time Markov chains. 
 CTQW of two particles over a  graph composed by $N$ sites
takes place in the  Hilbert space spanned by the orthonormal set of
position vectors $\{\ket{j,k}\}$  that describe the state in which one
particle is localized on site $j$ and the other on site $k$ with 
$j,k =1,\dots N$.  The dynamics of a  CTQW of two indistinguishable and
interacting particles over a 
homogeneous one-dimensional lattice, such as the line, is described by the total Hamiltonian:
\begin{align}
H_2&=H_0+H_{int}\label{H}\\
H_0 &= H_{1}\otimes\mathbb{I}+\mathbb{I}\otimes H_{1}\\
H_{int} &=U(|j-k|)\sum_{j,k=1}^N \vert j,k\rangle\langle j,k\vert,
\end{align}
 where $H_{1}= \epsilon\mathbb{I}-J\sum_j( \vert j\rangle\langle j+1\vert + \vert j+1\rangle\langle j\vert)$ 
 describes the hopping of a single particle between next-neighbors sites, $ H_{int}$ accounts for the interaction between the two particles, and $ U(|j-k|)$ shapes the strength of the interaction according to the distance between the pair, that, in the present case, is chosen to be:
 \begin{equation}\
U(|j-k|)= \left\lbrace \begin{array}{ccc} U \quad &\text{if}\quad & j=k \qquad \\
		U/3 \quad &\text{if}\quad & j=k+1 
\end{array}			
\right.			.				 								
\end{equation} 
As initial condition of the CTQW, we consider a state in which each particle is localized over a different site: 
\begin{equation}
|\Psi^{\pm}_0\rangle = \frac{1}{\sqrt{2}} (| j,k \rangle \pm |k,j \rangle) \quad \text{with}\quad j\neq k.
\label{psi0}
\end{equation} 
The symmetry of the initial state, i.e. the sign in Eq. \eqref{psi0}, then
determines weather the particles are bosons or  fermions, since the Hamiltonian $H$ 
conserves the symmetry of the state during its evolution. 
 By applying the unitary evolution $\Lambda(t)=\exp(-iHt)$ to the initial state  $|\Psi^{\pm}_0\rangle $ one obtains the dynamics of the pair. Notice that we set $\hbar=1$.
 \\
\indent Due to unavoidable interaction with the environment, noise is always present in realistic implementations of quantum walks. Some authors consider the possibility of lattice imperfections in the form of missing links, thus obtaining a percolation graph \citep{leung10,darazs13,rigovacca16}.  In order to simulate dynamical noise, we instead introduce a stochastic time-dependent term in the hopping Hamiltonian $H_{1}$, that randomizes the tunneling amplitudes between adjacent sites. Although fluctuating, transition amplitudes retain a finite value throughout the evolution. 
The single particle Hamiltonian  thus becomes a time-dependent random matrix $H_{1r}(t)$,   
written as the sum of the unperturbed term $H_1$ and a stochastic  contribution affecting the transition rates between adjacent sites
\cite{benedetti16}:
\begin{equation}
 H_{1r}(t)= H_{1} + J\sum_j g_j(t)( \vert j\rangle\langle j+1\vert + \vert j+1\rangle\langle j\vert),
\end{equation}
where  
the coefficients $\lbrace g_j(t)\rbrace$ are the time-dependent fluctuations of the tunneling amplitude that introduce decoherence in our description of CTQWs,
and the two-particle Hamiltonian reads:
\begin{equation}\label{Htot}
H_{2r}(t)=H_{1r}(t) \otimes \mathbb{I} + \mathbb{I} \otimes H_{1r}(t) + H_{int}.
\end{equation}
 Each $g_j(t)$  is independent from the others and is modeled as  random
 telegraphic noise (RTN), i.e. as a dichotomic variable which can  only
 jump between two values $g_j(t)=\pm g_0$, with a certain switching rate
 $\xi$. Other authors use RTN is characterized by an exponentially decaying
 autocorrelation function: \begin{equation}
C(t)=\langle g_j(t)g_k(0)\rangle =\delta_{jk}\, g_0^2 \,e^{-2\xi t},
\label{autoRTN}
\end{equation}
where the Kronecker delta $\delta_{jk}$ expresses 
the fact that the random processes $\{g_j\}$ are independent 
each others.  Starting from the
initial state $\rho_0=\ketbra{\Psi^{\pm}_0}{\Psi^{\pm}_0}$, the dynamics
of the two particles for a single realization of the stochastic
processes $\{g_j(t)\}$ is governed by the evolution operator 
\begin{equation}
 \Lambda(t)=\mathcal{T} \exp\left(-i \int_0^t H_{2r}(s)\, ds \right),
\end{equation}
where $\mathcal{T}$ is the time-ordering operator.  The time-evolution
of the two-particle CTQW is thus calculated by  averaging  the single
realization dynamics $\Lambda(t)\rho_0\Lambda^{\dagger}(t)$, over all
the possible realizations of the stochastic processes:
\begin{equation}\label{dinamica}
\rho(t)=\left\langle \Lambda(t)	\rho_0 \Lambda^{\dagger}(t)\right\rangle_{\lbrace g_j(t)\rbrace}.
\end{equation}
Without loss of generality, the dynamical parameters 
may be rescaled in terms of the hopping strength $J$. From now on 
we will describe the dynamics in terms of a dimensionless time 
and switching rate: \begin{equation}
t \rightarrow Jt \equiv\tau \qquad \xi\rightarrow \xi/J\equiv \gamma.
\end{equation}
Upon looking at the autocorrelation function of the RTN 
in Eq. \eqref{autoRTN}, one may distinguish two regimes, which 
characterize different time scales for the noise. If $\gamma\gg1$, 
we talk about {\it fast} noise, because this situation 
corresponds to the two particles evolving in a fast fluctuating 
environment, i.e. where the bistable fluctuators $g_j(t)$ 
flip according to a very large switching  rate. On the contrary, 
the {\it slow} noise regime arises when $\gamma\ll1$, and it describes 
the case of quasi-static disorder \cite{benedetti12i,benedetti16}.
 %%%%%%%%%%%%%%%%%%%%%%%%%%%%%%%%%%%%%%%%%%%%%%%%%%%%%%%%%%%%%%%%%%
\begin{figure}[t!]
\includegraphics[width=0.97\columnwidth]{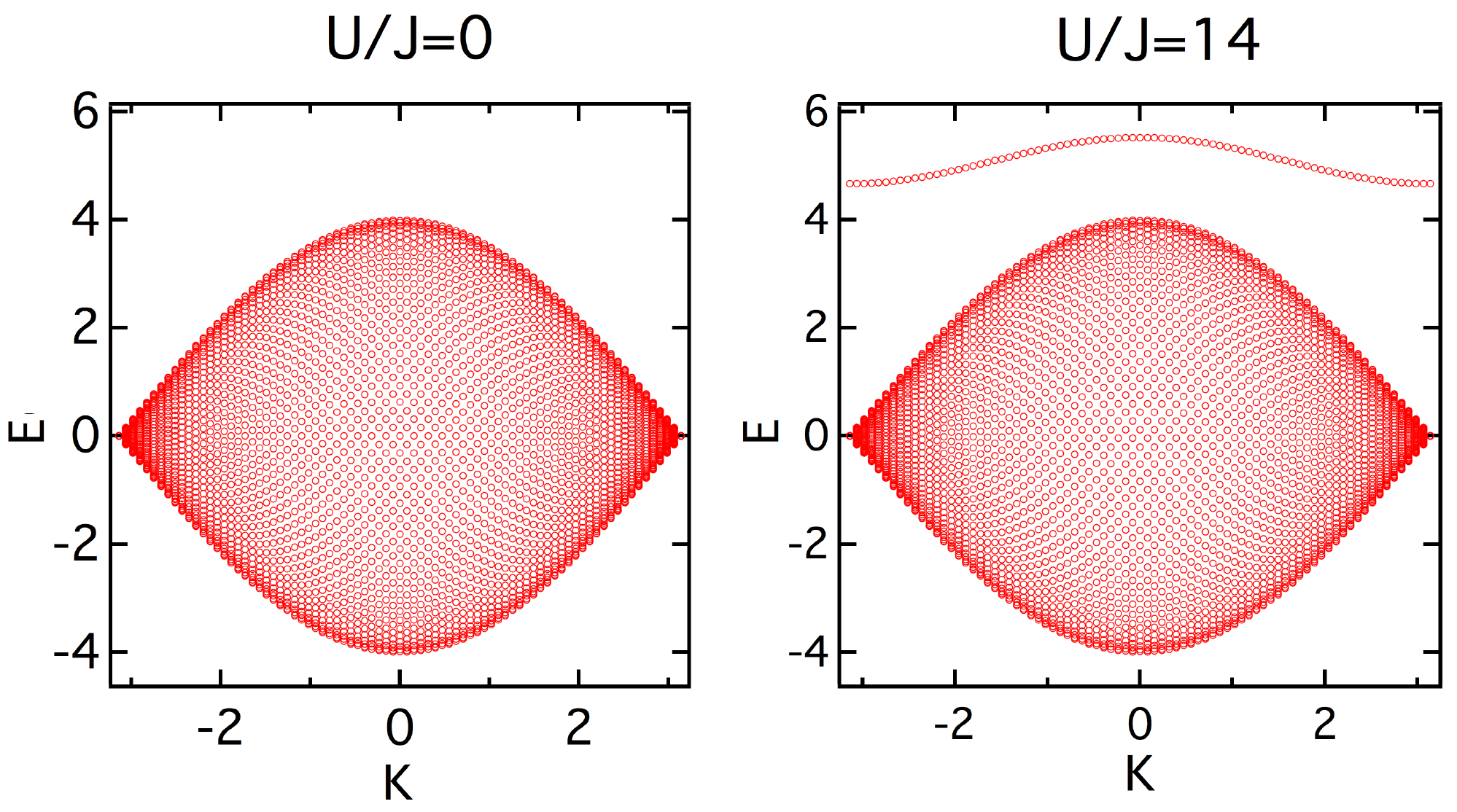}
\caption{(color online) 
Band-structure of the Hamiltonian $H_2$ in Eq. \eqref{H} for the case of
two fermions on a lattice of $N=80$ sites. The left plot corresponds 
to $U/J=0$ and the right one to $U/J=14$.} \label{fig1}
\end{figure}
%%%%%%%%%%%%%%%%%%%%%%%%%%%%%%%%%%%%%%%%%%%%%%%%%%%%%%%%%%%%%%%%%%%%%%%%%%%%
\section{Quantum walk of two interacting particles}
\label{sec3}
In order to investigate the propagation of two interacting particles in
the presence of dynamical noise, we first review the case of a noiseless
evolution in subsection A, then we analyze the effect of decoherence on
the two-walker dynamics in subsection B. 
\par
The two particles are initially localized in two sites in the middle of
a 1D lattice with periodic boundary conditions  containing $N=80$ sites.
As noise is introduced, the time evolution of the CTQW is computed by
numerically calculating the expression  of Eq.(\ref{dinamica}).  We
evaluate the ensemble average over 5000 different noise realizations.
Simulations are performed by implementing a specific GPU accelerated
code. In this way the simulation time  (for $N=80$ sites and 5000 noise
realizations) is limited to approximately 25 minutes on an NVIDIA
Tesla K40 board.  
\par
For each noise realization, the switching times, i.e. the times 
at which the stochastic processes $\{g_j\}$ jump from one value to the
other thus changing the transition amplitudes between $J\pm g_j(t)$
are generated by using the Monte Carlo method. To maximize the effect of
noise, we set the noise amplitude to $g_0=0.9J$. A more
comprehensive analysis on the effects of the noise amplitudes may be
found in  \citep{benedetti16}. In all the simulations, the evolution is
stopped before the particles may reach the lattice boundaries. This is
equivalent to study the dynamics on an infinite lattice, and allows us
to isolate the effect of the noise from the possible interference
effects due to boundary conditions.
\par 
By tracing out one particle, we may 
characterize the propagation in terms of single particle variance 
\begin{equation}
\sigma^2(t)= \sum_x\langle x^2(t) \rangle -\langle x(t) \rangle^2\,,
\end{equation} 
with $\langle x^k\rangle= \sum_i i^k \rho_{ii}^1(t)$ and $\rho^1(t)$ 
is the single-particle reduced density matrix obtained by tracing out
the other particle.  This quantity is meaningful to evaluate the quantum
walk spread in time and, consequently, to observe the transition from
ballistic ($\sigma^2\propto t^2$) to diffusive ($\sigma^2\propto t$)
propagation. Furthermore, we evaluate the occupation number of the
lattice sites during the evolution \begin{equation}
\langle n_k(t)\rangle = 2\sum_j\rho_{kj,kj}(t)\,,
\end{equation}
where $\rho_{kj,kj}(t)$ are the populations of the two-particle density
matrix in the $\{\ket{k,j}\}$ basis. This quantity represents the
average number of particle in each site; it provides information about
the spatial distribution and about the localization of the pair.  
\par
As the results obtained for bosons and fermions are qualitatively
identical, for the sake of brevity hereafter we will present results
relative to
the case of two interacting fermions.
\par
In all the figures quantities reported are dimensionless.  

%%%%%%%%%%%%%%%%%
\begin{figure}[hptb]
\includegraphics[width=0.96\columnwidth]{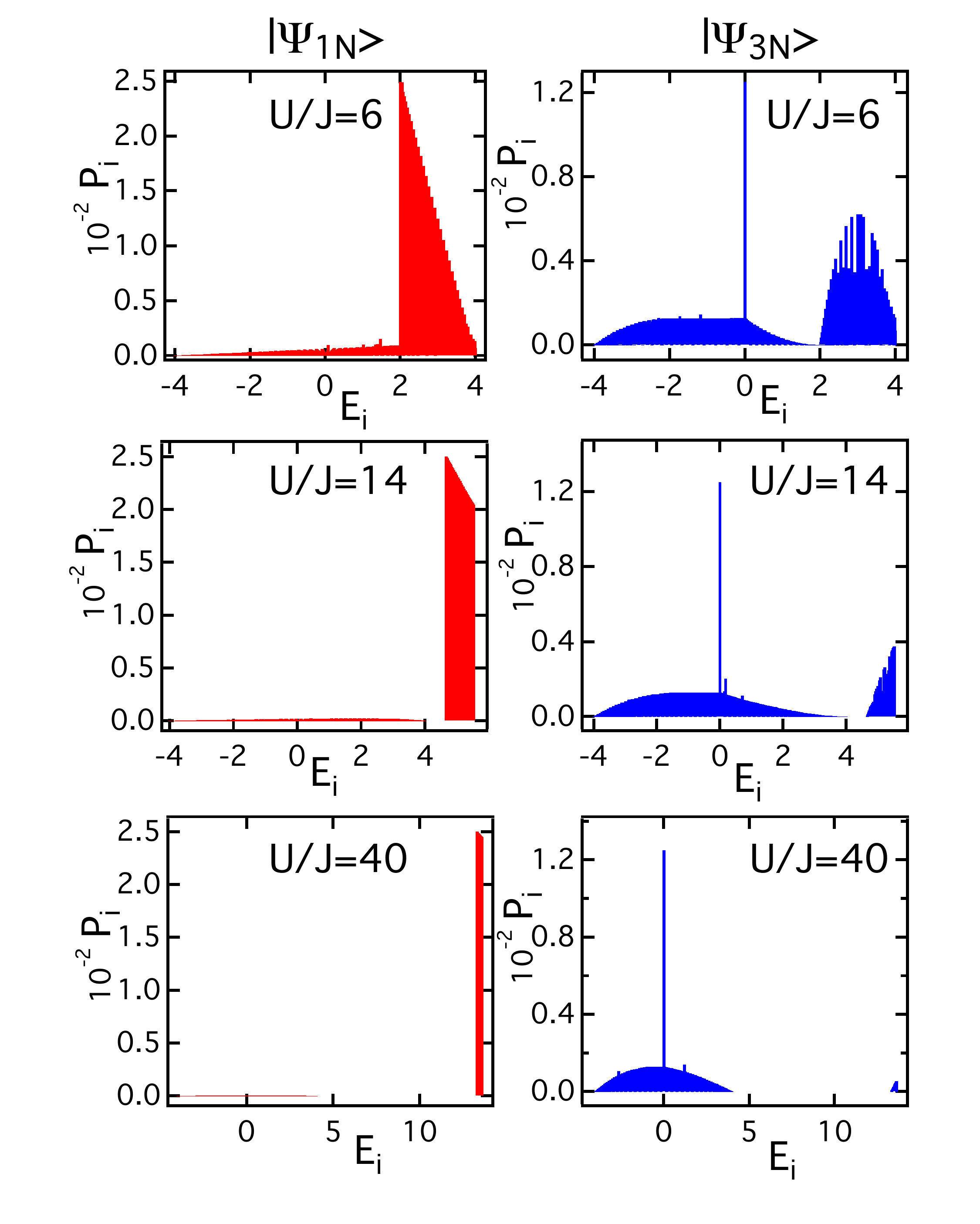}
\caption{(Color online) The left column of plots 
show the projections $P_i$ of the fermionic state 
$|\Psi_{1N}\rangle$ (corresponding to particles 
initially located on next-neighbors sites) 
on the eigenstates of the unperturbed Hamiltonian 
$H_2$ of Eq.\eqref{H} ($E_i$ is the eigenvalue corresponding to 
the $i$-th eigenvector).
Different plots refer to different
 interaction strengths. The right columns of plots show the
 corresponding projections for the
 fermionic state $|\Psi_{3N} \rangle$ (particles initially
 located on 3-neighbors sites).}
\label{fig2}
\end{figure}
%%%%%%%%%%%%%%%%%%%%%%%%%%%%%%%%%%%%%%%%%%%%%%%%%%%%%%%%%%%%%%%%%%%%%%%%%%%%
\subsection{Noiseless dynamics}
\label{noiseless}
In the absence of noise, the Hamiltonian  $H_2$ in Eq. \eqref{H} is
obtained by adding a term containing the discretization of the Laplacian
operator ($H_{1}$) plus an interaction term ($H_{int}$) depending only
on the relative distance among particles. In a homogeneous lattice,
where the hopping amplitudes are all equal to $J$, the translational
invariance allows one to solve the Schr\"{o}dinger equation by the
ansatz $\Psi(j,k)=\mathrm{exp}(iKR)\phi_{K}(r) $, where we introduced
the center of mass of the pair $R=(j+k)/2$ and the relative
inter-particle distance $r=|j-k|$.  In this picture, $K$ is the
quasimomentum and it assumes only discrete values due to the lattice
finite size, $K=2\pi\nu/N$ with $\nu=1,2,\dots,N$, $\phi_{K}(r)$ is
the pair wave function. The band-structure of $H_2$, whose explicit
derivation can be found in \cite{scott94,valiente08}, represents an
essential ingredient to study the two-particle dynamics. As displayed in
Fig. \ref{fig1}, in the absence of interaction ($U=0$), the
band-structure consists of a unique band identical for bosons and
fermions. For finite next-neighbors interaction strength, one observes
the formation of a small band, called \textit{miniband}, whose energy at
the edge of the first Brillouin zone ($K=\pi$) is given approximately by
$U/(3J)$; in the case of bosons, the additional on-site interaction
generates a second miniband at higher energies, $U/J$.  All the
remaining states are contained in the \textit{main band}, that ranges
approximately from $-4J$ to $4J$ \cite{valiente08,nguenang09,boschi14}.
It is worth noting that the appearance of a single miniband is related
to the form of the interaction we have chosen. If second neighbors were
affected by interaction, one would observe the formation of a second
band. On the other hand, the symmetry with respect to the energy depends
on the hopping range, thus more complicated and not symmetric band
structures are obtained when long-range hopping terms are included
\citep{chattaraj2016}.
\par
%%%%%%%%%%%%%%%%%%%%%%%%%%%%%%%%%%%%%%%%%%%%%%%%%%%%%%%%%%%%%%%%%%
\begin{figure}[ht!]
\includegraphics[width=0.96\columnwidth]{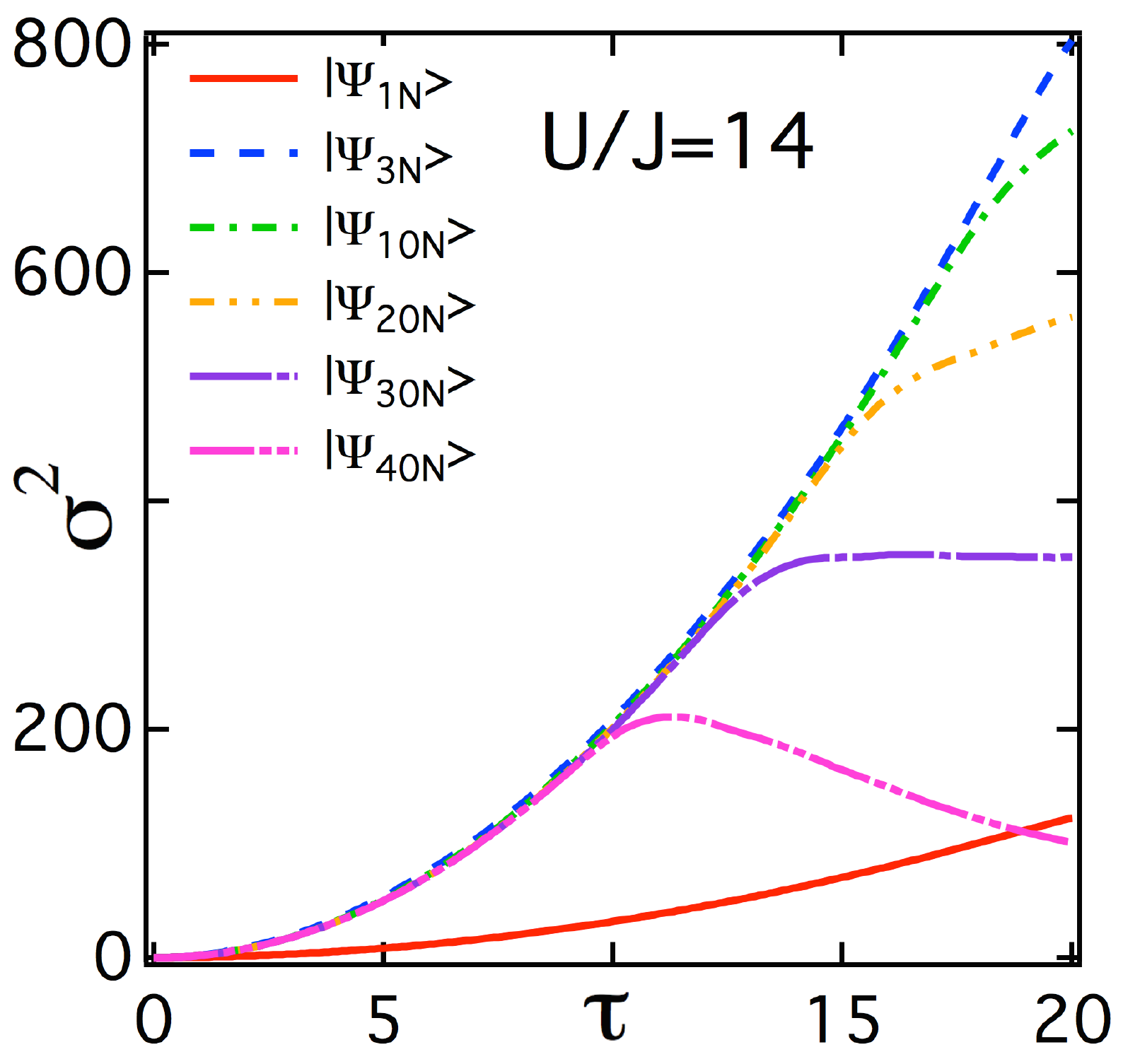}
\caption{(color online) 
Single particle variance $\sigma^2(t)$ as function of time for two
fermions with different starting sites: next-neighbors (red solid line),
3-neighbors (blue dash), 10-neighbors (green dash-dot), 20-neighbors
(yellow dash-two dots), 30-neighbors (violet long dash-dash) and
40-neighbors (pink long dash-two dashes). The interaction is here fixed
at $U/J=14$. In order to compare the different dynamics the initial
value is set to $\sigma(\tau=0)=0$ for all the curves. Since the farther
are the particles in the initial condition, the sooner they reach the
boundaries, we observe changes in the slope of the various curves (for
initial particle distance larger than 10 sites) due to spurious
interference effects.} \label{fig3}
\end{figure}
%%%%%%%%%%%%%%%%%%%%%%%%%%%%%%%%%%%%%%%%%%%%%%%
\par
The dynamics of the two walkers strongly depends on the initial state.
If one considers a suitble interaction regime $(U/J>6)$, particles
initially occupying next-neighboring sites give raise to a bound
pair \cite{winkler06} that propagates as one single packet 
through the lattice, a phenomenon called \textit{cowalking dynamics}, 
independently on the sign of the interaction. Usually the eigenstates 
of the minibands are associated with such \textit{bound states}. 
Conversely, eigenstates belonging to the main band, the so called
\textit{scattering states}, are characterized by a delocalized 
wavefunction.  In other words, the dynamics of fermions starting 
from next-neighboring sites is mainly confined to the miniband, 
whereas particles starting from $n$-neighboring sites, with $n>1$, 
belong to the main band. This feature is more evident when the 
interaction strength increases, as in this case there is a proper 
gap between the two bands, whose central width is given by 
$\Delta_{K=0}=U/3 + 12/U - 4$ in the limit of $N\rightarrow\infty$,
\cite{scott94}.  
\par
In order to analyze quantitatively this effect, we consider the
projections 
$$P_m=\sum_k |\langle \bar{\Psi}_{m;k} | \Psi_0\rangle |^2$$ 
of a state $ |\Psi_0\rangle $ onto the $m-th$ eigenstantes $
|\bar{\Psi}_{m;k}\rangle $ of the Hamiltonian $H_2$, where the
summation over $k$ accounts for the degeneracy, and evaluate 
this quantity for different value of the interaction strengths.  
In Fig. \ref{fig2}, we compare the projections
$P_m$ for particles starting from next-neighbors sites 
\begin{equation}
\ket{\Psi_0}=|\Psi_{1N} \rangle = 
1/\sqrt{2}(|j_0,j_0+1\rangle-|j_0+1,j_0\rangle)
\end{equation}
to the ones initialized in 3-neighbors sites 
\begin{equation}
\ket{\Psi_0}=\ket{\Psi_{3N} }= 
1/\sqrt{2}(|j_0,j_0+3\rangle-|j_0+3,j_0\rangle)\,.
\end{equation}
For $ U/J=6 $ projections are distributed on both bands since the energy
levels form a quasi-continuum, i.e. the two sub-bands are not completely
detached. The support of $\ket{\Psi_{1N}}$ is mostly on the miniband, 
but still there are some contributions from the scattering band.
The converse happens for $\ket{\Psi_{3N}}$, and the fact that its 
projections are  smaller than those of a bound state is due to the much
larger number of states belonging to the main band. The larger is the
interaction, the smaller are the projections of a bound (scattering) 
state onto the main (mini-) band. Since a proper gap separates the two 
bands, two distinct dynamical regimes arise, each one being a
characteristic feature of a definite sub-band.
\par
Such distinction becomes apparent if one compares the
single-particle variance for states initially localized in next-, 3-,
10-, 20-, 30-, 40-neighbors sites, see Fig. \ref{fig3}. Except for a
shift factor depending on the relative coordinate $r$, scattering states
exhibit the same ballistic propagation ($\sigma^2\propto t^2$). Bound
states $\ket{\Psi_{1N}}$ are still characterized by a parabolic profile,
but the slope is reduced due to the interaction range. It is worth to
remember that the particle velocity is given by the slope of the band;
since the slope of the scattering band is always larger than the one of
the miniband - and it does not vary with the interaction strength - ,
states with a large number of components in the scattering bands have
faster velocity components thus achieving a faster ballistic
propagation. On the other hand, the slope of the mini-band gets reduced
with increasing interaction, thus the spread velocity tends to zero in
the limit of infinite $U$. In this regime the variance will be frozen in
its initial value ($\sigma = 0.25$) and the particles will be localized
on the starting sites by the strong next-neighbors interaction.  
%%%%%%%%%%%%%%%%%%%%%%%%%%%%%%%%%%%%%%%%%%%%%%%%%%%%%%%%%%%%%%%%%%
\begin{figure}[t]
\includegraphics[width=0.98\columnwidth]{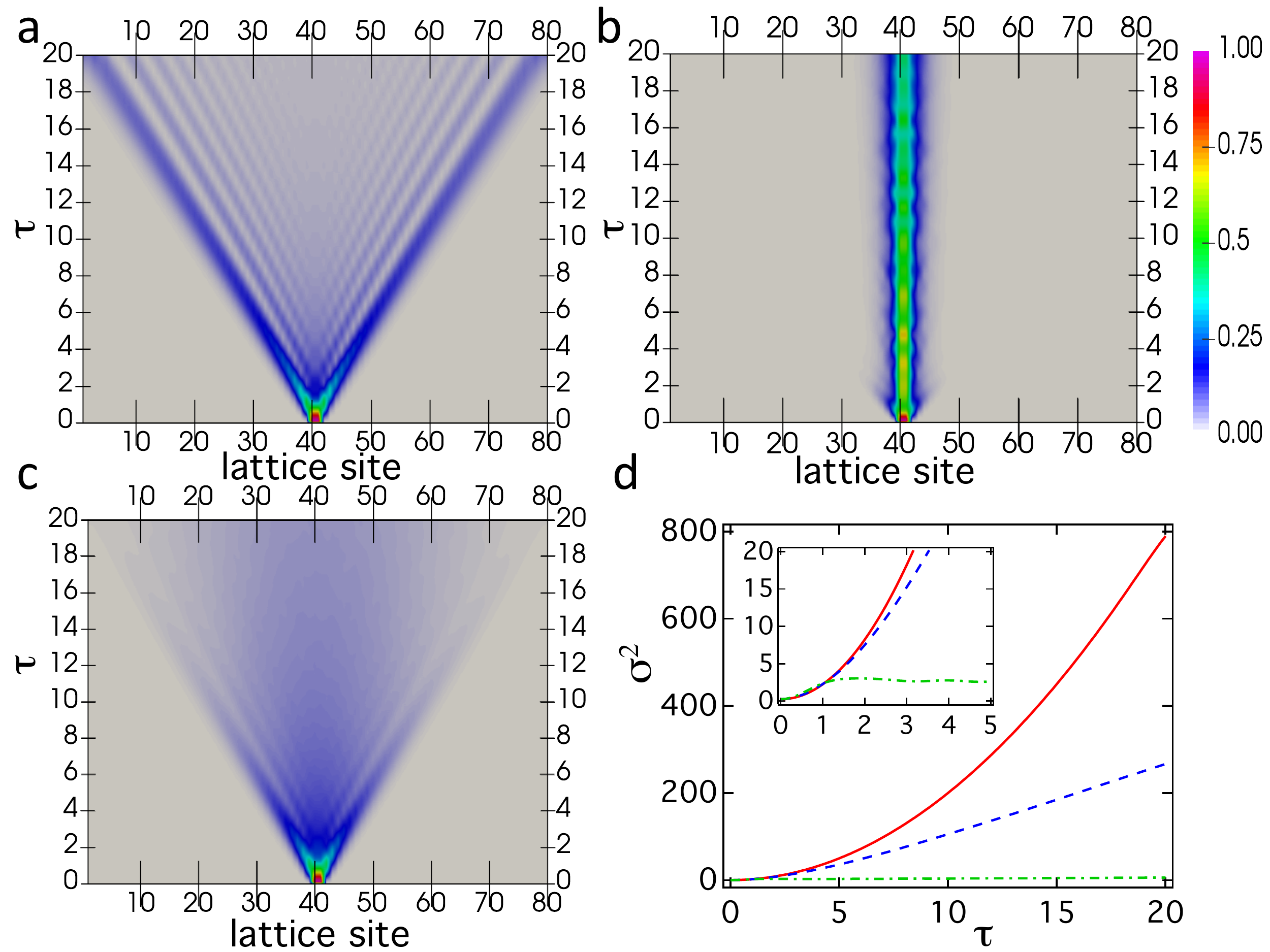}
\caption{(color online) 
Occupation number maps as a function of time and lattice sites, $\langle
n(t)\rangle$, for two non-interacting ($U/J=0$) fermions initially
located on next-neighbors sites. The unitary dynamics (a) is compared
with the fast noise dynamics (c) with switching time $\gamma=10$ and
slow noise propagation (b) with switching time $\gamma=0.01 $. Noise
amplitude set to $g_0=0.9$; (d) Single-particle variance of the position
as function of time in the absence of interaction $U=0$; different
curves account for the noiseless case (red solid line) and for different
values of the switching time - $\gamma=0.01$ (slow RTN, green dash-dot),
$\gamma=10$ (fast RTN, blue dash), whereas the noise amplitude is set to
$g_0=0.9$.The curve is identical for both bosons and fermions. Inset:
zoom on the first part of the dynamics to highlight the localization
induced by the slow noise.} \label{fig4}
\end{figure}
%%%%%%%%%%%%%%%%%%%%%%%%%%%%%%%%%%%%%%%%%%%%%%%%%
\begin{figure}[hptb]
\includegraphics[width=0.96\columnwidth]{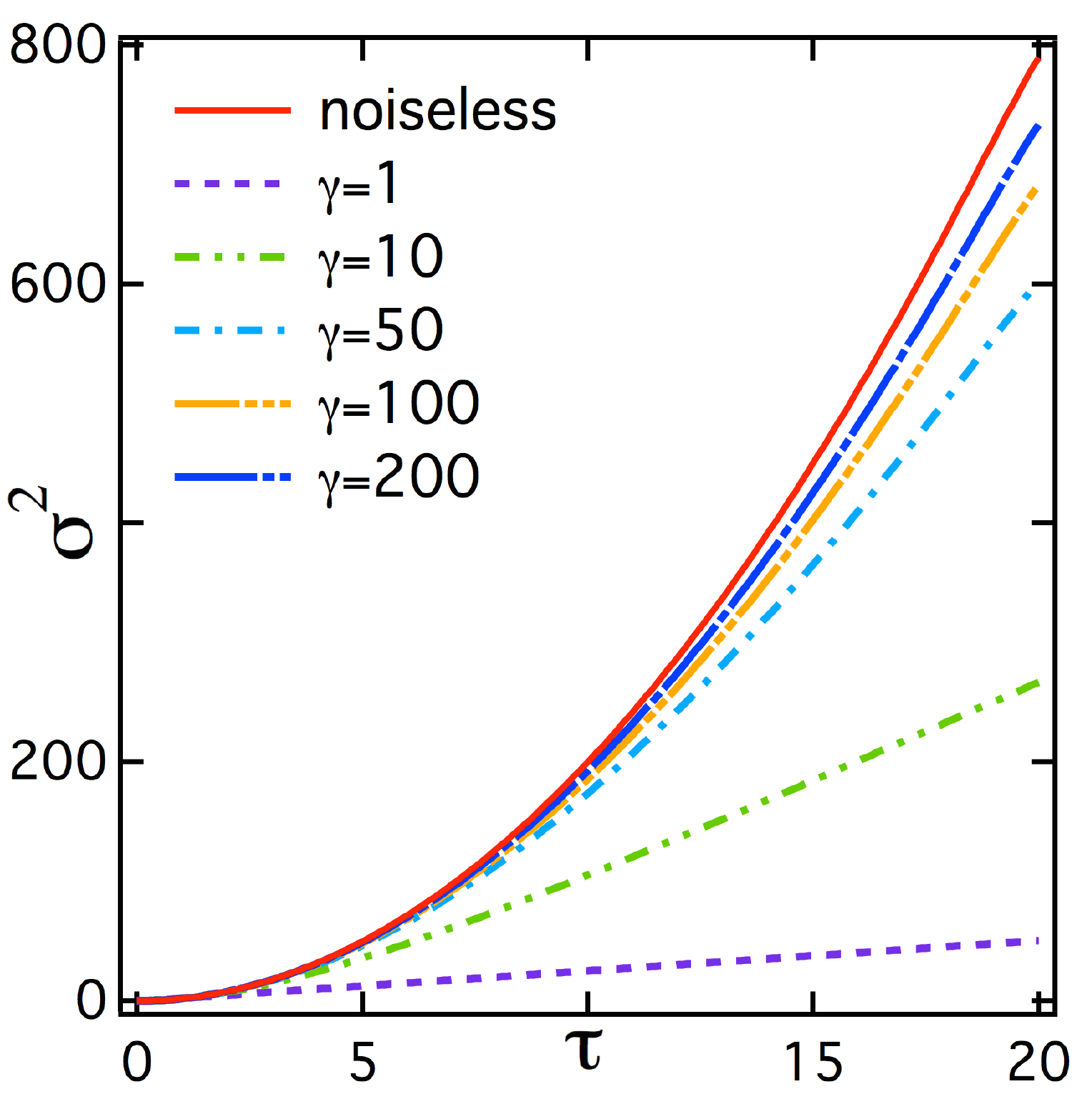}
\caption{(color online) 
 Single particle variance $\sigma^2(t)$ over the lattice as a function of time for two non-interacting fermions in the fast noise regime for different switching rates. Each curve refers to a different $\gamma$ value: $\gamma=0$ (red solid line), $\gamma=1$ (violet dash), $\gamma=10$ (green dash-two dots), $\gamma=50$ (light blue dash-dot), $\gamma=100$ (yellow long dash-dot) and $\gamma=200$ (blue long dash-two dots). Noise amplitude is set to $g_0=0.9$.
}
\label{fig5}
\end{figure}
%%%%%%%%%%%%%%%%%%%%%%%%%%%%%%%%%%%%%%%%%%%%%%%%%
\subsection{Noisy dynamics}
\label{noisy}
In order to understand how dynamical noise affects the propagation of
the two particles, we first analyze the case of two indistinguishable
non-interacting particles, where the exchange symmetry is the only
ingredient added with respect to the single particle picture
\citep{benedetti16}.  Similarly to single-walker case, fast noise drives
a transition from quantum ballistic to classical diffusive propagation.
This feature is evident when considering the single particle spread
$\sigma^2(t)$ over the lattice. This is shown in Fig. \ref{fig4}(d),
where the variance is quadratic at short times while at longer times it
grows linearly.  In the fast noise regime $\gamma\gg1$, the slope of the
variance decreases with smaller switching rates, whereas for larger
values of $\gamma$ one recovers the ballistic dynamics, 
as is evident from Fig. \ref{fig5}. 
 \par
In the very fast noise regime, indeed, the time scales of the system and
the environment are well separated, and the  oscillations in the
transition amplitudes happen so fast with respect to the walkers
dynamics that they no longer affect the evolution of the particles.  By
comparing the evolution of the occupation number $\langle n_j\rangle$
in the noiseless scenario with the case of fast noise,  see Fig.
\ref{fig4}(a-c), one observes that noise wipes out the clear
interference pattern, but preserves the antibunching behavior typical of
fermions.  It is worth noting that in both the noiseless and fast noise
cases the particles reach the boundaries of the lattice at the same time
even if they are differently distributed among the  sites.  Indeed, in
Fig. \ref{fig4}(a), the interference pattern has a larger intensity in
the farthest points from the origin, while fast noise tends to focus the
wavefunction amplitude in a central area around the initial positions.
In the same way, fast noise rearranges the weight distribution among the
velocity components thus affecting the average velocity, as detected by
the single particle variance. In the previous subsection we have shown
that the noiseless dynamics is completely determined by the
band-structure. 
\par
As the translational invariance is broken by random dynamical 
noise, it is no longer possible to define a proper band-structure. 
On the other hand, the dynamics of the walkers subject to very fast 
noise  approaches the noiseless dynamics. Therefore, we may see fast
noise as a small perturbation to the noiseless case and, to a first
order approximation, still analyze the dynamics in terms of bands. 
In turn, in the presence of slow noise, the features described 
above are absent. The propagation is suppressed, as one can see 
in Fig. \ref{fig4}(b), and the occupation number during the evolution
is different from zero only for few sites close to the initial
positions. Moreover, after a certain time evolution, whose value depends 
on the switching time of the noise, the variance achieves a saturation 
value, see Fig. \ref{fig4}(d), and the system undergoes an 
Anderson-like dynamical localization phenomenon.
%%%%%%%%%%%%%%%%%%%%%%%%%%%%%%%%%%%%%%%%%%%%%%
\begin{figure}[t!]
\includegraphics[width=0.96\columnwidth]{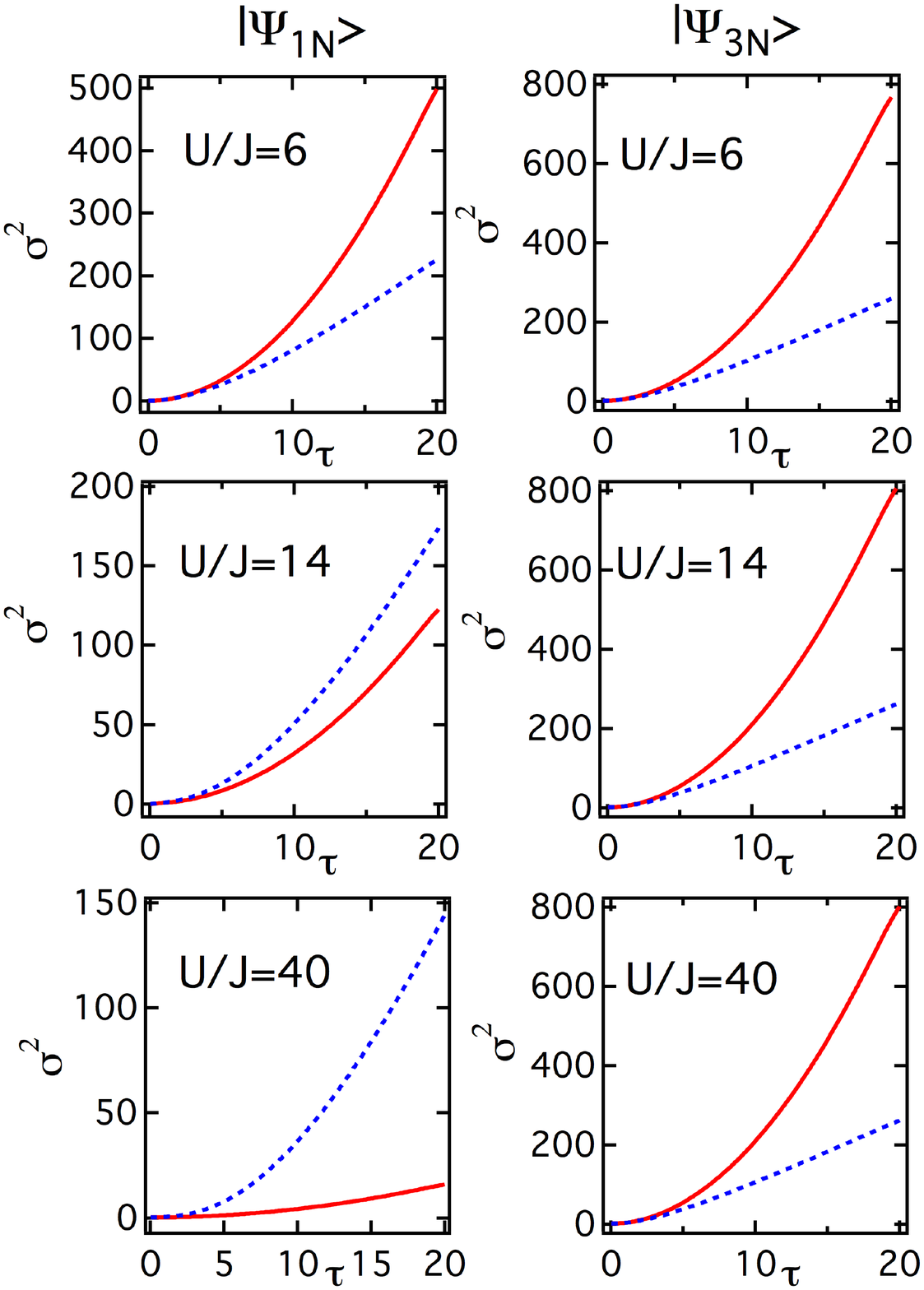}
\caption{(color online) 
 Single particle variance $\sigma^2(t)$  as a function of time for two
 fermions starting from next-neighbors sites $|\Psi_{1N}\rangle$ (left
 column) and third-neighbors $|\Psi_{3N}\rangle$ (right column): each
 panel considers a different interaction strength $U/J$, and compares
 the noiseless evolution (solid red line) with the one in fast noise
 regime (dotted blue line), whose amplitude and switching time are
 respectively $g_0=0.9$ and $\gamma=10.0$.  }
\label{fig6}
\end{figure}
%%%%%%%%%%%%%%%%%%%%%%%%%%%%%%%%%%%%%%%%%%%%%%%%%
\begin{figure}[ht]
\includegraphics[width=0.96\columnwidth]{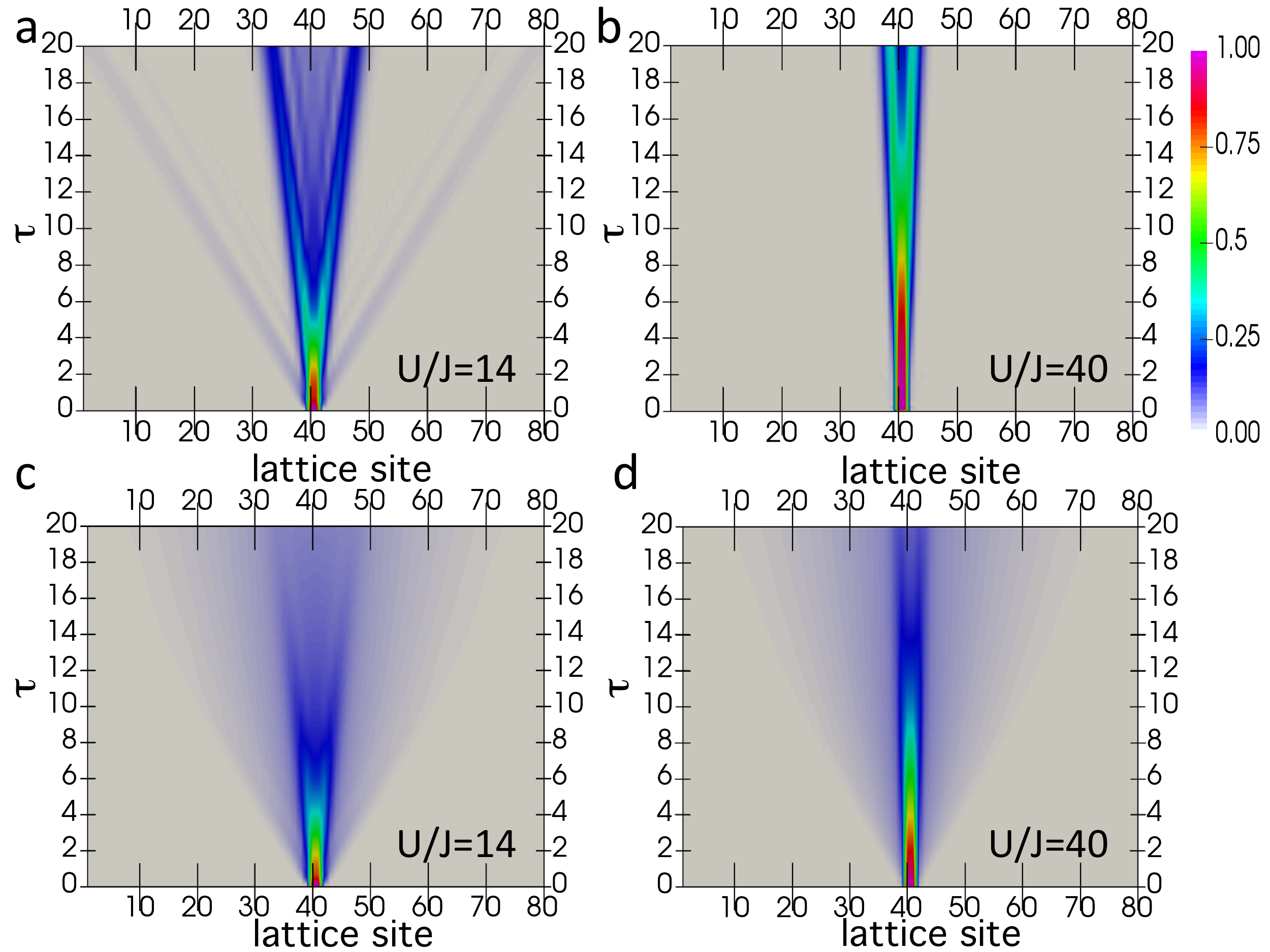}
\caption{(color online) 
Occupation number maps as a function of time and lattice site, $\langle
n_j(t)\rangle$, for two fermions initially located on next-neighbors
sites. Different interaction strengths are considered: $U/J=14$ (left
column) and $U/J=40$(right column). By rows, the unitary dynamics (top)
is compared with the fast noise dynamics (bottom) with switching time
$\gamma=10$ (fast RTN) and noise amplitude set to $g_0=0.9$.  }
\label{fig7}
\end{figure}
%%%%%%%%%%%%%%%%%%%%%%%%%%%%%%%%%%%%%%%%%%%%%%
\par
Let us now move attention to interacting particles. In Fig.  \ref{fig6}
we compare the single particle variance in the noiseless case to the
fast noise one, for different values of the interaction strengths. If we
consider particles starting from next neighboring sites, i.e. the
initial state is $|\Psi_{1N}\rangle$, we see that if the ratio $U/J$ is
increased.  the variance in the presence of fast noise becomes larger
compared to the noiseless case.  Taking into account that whenever
$U/J>6$, a  gap between the two sub-bands appears, see Fig. \ref{fig1},
it becomes apparent that noise provides access to a novel regime where
the particles acquire faster propagation components, which show up in
the single particle
variance. 
\par
As we have already observed before, upon excluding few components with
small weights on the main band, the noiseless evolution of
$|\Psi_{1N}\rangle$ is mainly confined to the miniband, see Fig.
\ref{fig2}, where each $K$ component propagates with a smaller velocity
with respect to the the scattering band.  Fast noise regime appears to
preserve the band structure and allows for a redistribution of the
wavefunction component between the two sub-bands, thus enabling a faster
propagation through the lattice.  This feature is more evident for
larger values of the interaction, as projections on the main band
becomes smaller and the noiseless dynamics slower. Clearly such gain
effect does not show up when the initial state has most of its
components on the main band, as in the case of $|\Psi_{3N}\rangle$, see
second column in Fig.  \ref{fig6}. In this case, the redistribution of
the wavefunction  brings components into the miniband and does not allow
for a faster dynamics. Indeed, the noiseless variance is always faster
than the noisy one and, consinstently with the results of previous
section, this phenomenon 
is independent on the value of the interaction strength.  
\par
This behavior is captured also by the occupation number, shown in Fig.
\ref{fig7}.  Here we observe that for large interaction values
($U/J=14,40$), see Fig. \ref{fig7}(c,d), new  areas of the lattice  are
accessible to the walkers in the presence of fast noise compared to the
unperturbed case Such a broadening in the spatial distribution of the
pair comes from faster velocity components that allow for propagation
even if the interaction would be strong enough to induce localization,
see Fig. \ref{fig7}(b).  
\par
\begin{figure}[t!]
\includegraphics[width=0.96\columnwidth]{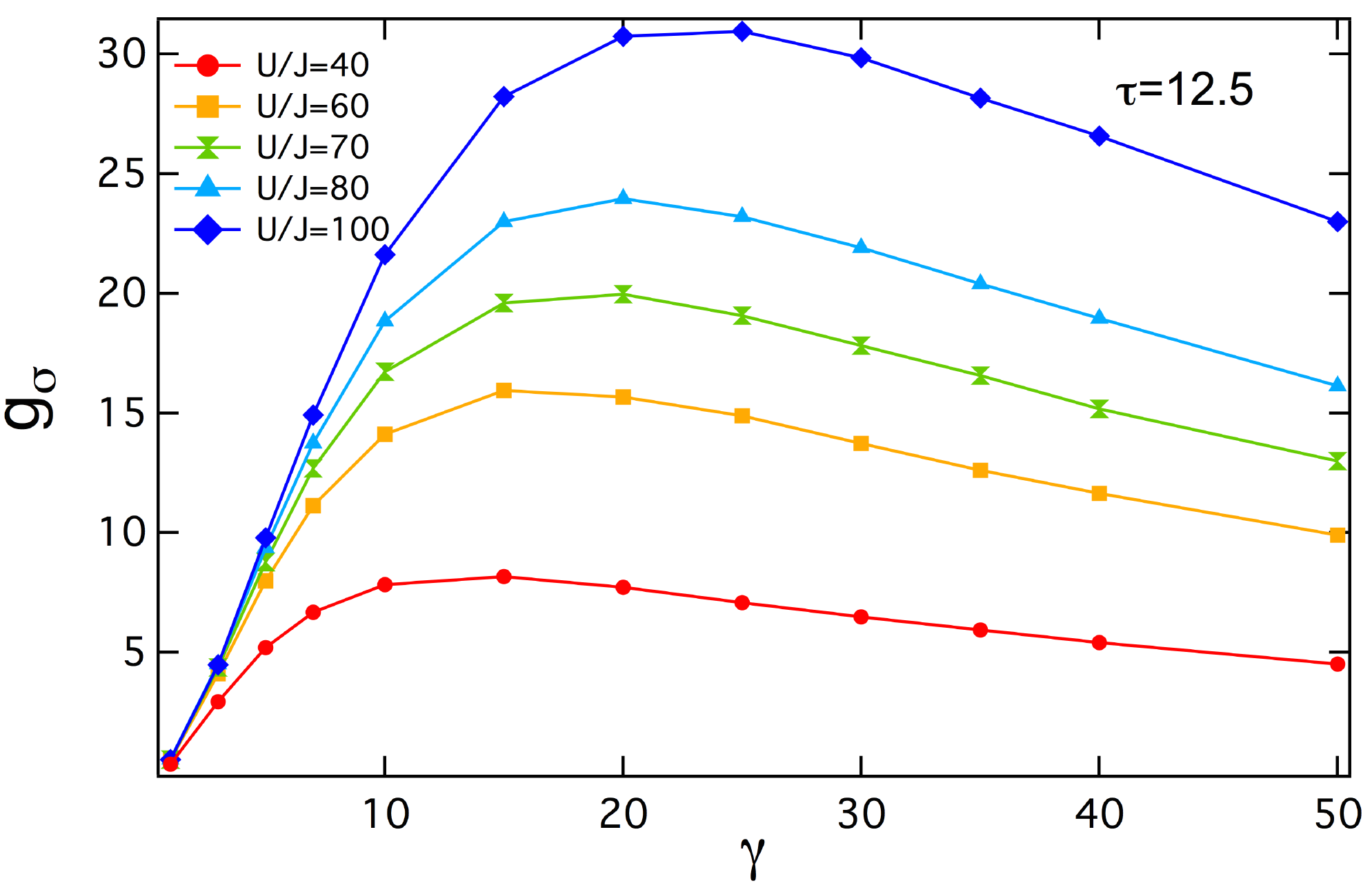}
\caption{(color online) Single particle variance gain $g_{\sigma}$ for
two fermions, initially located in next-neighbors sites, as a function
of the switching rate parameter $\gamma$, with $g_0=0.9$. The variances 
are calculated at $\tau=12.5$, when the dynamics is not yet affected by
the boundary conditions. Each curve corresponds to a different 
value of the interaction strength, we have $U/J=40$ (red circles), 
$60$ (yellow squares), $70$ (green hourglasses), $80$ (light blue 
triangles), and $100$ (blue rhombus).}
\label{fig8}
\end{figure}

Finally, as a further evidence supporting our conjecture that noise
allows for faster wavefunction components, we investigate the interplay
between the parameter $\gamma$ and the interband gap $\Delta$. In
particular, in order to see whether the observed faster propagation 
in the noisy regime is linked to the characteristic parameters of 
the noise and  the system, we introduce the variance gain 
\begin{equation}
g_\sigma= \sigma_{fast}^2/\sigma_{nonoise}^2 -1
\end{equation}
and analyze its beahviour a function of $\gamma$ at a fixed time 
in the evolution. In particular, in Fig. \ref{fig8} we show the 
results for $\tau=12.5$. Each curve corresponds to a different 
interaction strength. In all cases, the variance gain displays a
similar behavior: the gain increases with increasing $\gamma$ up to a
maximum value after which it decreases and, in the limit of $\gamma
\rightarrow \infty$, it vanishes, in agreement with the results
shown before about very fast fluctuations. It turns out that each peak
shifts to larger gamma values for larger gaps (i.e. larger interaction
values), which means that a larger gap needs a faster noise to maximize
the gain. By repeating the same calculations for different times during
the dynamics, we find out that the optimal $\gamma$ value corresponding
to the peak do not change much with time, while the value of the 
maximum gain changes. 
 %%%%%%%%%%%%%%%%%%%%%%%%%%%%%%%%%%%%%%%%%%%%%%%%%%%%%%%%%%%%%%%%%%
%%%%%%%%%%%%%%%%%%%%%%%%%%%%%%%%%%%%%%%%%%%%%%%%%%%%%%%%%%%%%%%%%%%%%%
\section{CONCLUSIONS}
\label{sec4}
Two-particle quantum walks are paradigmatic systems to address the
interplay between particles' indistinguishability and particles'
interaction, and to analyze in details the resulting propagation
regimes. Besides the fundamental interest,  two-particles quantum walks
are implemented on different platforms also with the aim of studying
multiple quantum interference and to simulate physical, chemical and
biological complex systems.  In turn, experimental realizations of QWs
may be subject to imperfections and defects, or to external
perturbations, and those different sources of noise may change, also
dramatically, the dynamical behaviour of the walkers. 
\par
In this paper, in order to analyze more realistic scenario for quantum
walks and  to explore new dynamical regimes for the particles, we have
addressed the decoherent dynamics of two indistinguishable and
interacting particles over one-dimensional lattices with random, 
time-dependent, tunneling amplitudes. In particular, the hopping 
amplitudes between adjacent sites of the lattice have been modeled as 
independent stochastic processes in the form of non-Gaussian random 
telegraphic noise. Depending on the value of the switching rate of 
the RTNs, and on the strength of the interaction between the walkers, 
different dynamical regimes arise.
\par
In order to compare our results to the noiseless case,  we have first
shown that the propagation of two interacting particles moving on a
perfect one-dimensional lattice is strongly affected by their initial
condition and the interaction strength. This feature gives rise to
different  dynamics that may be understood in terms of the band
structure of the systems: two sub-bands, in fact, appears and become
progressively more detached as the interaction between the two particles
increases.  We have then analyzed the effect of noise on the time
evolution of the walker, in terms of their position variances and the
occupation numbers over the lattice. 
\par
Our results suggest  that fast noise redistributes the wavefunction
components between the two sub-bands, giving rise to new dynamical
regimes that cannot appear without the contribution of noise.  Under
appropriate initial condition, the dynamics in the presence of fast
noise leads indeed to a faster propagation, as revealed by the single
particle variance, and a more spread in the spatial distribution, as
revealed by the occupation number. We have also analyzed the dependency
of the faster dynamics  on the characteristic parameters of both noise
and system, and have shown that  a larger band gap (originated from a
stronger interaction value) tipically needs a faster noise to maximize
the variance gain.  On the other hand, the propagation is suppressed in
the slow noise regime, where the system displays a dynamical {\em
Anderson-like} localization phenomenon, in agreement with previous
results \cite{benedetti16} for a single walker.
\par
Overall, our results show that upon tuning the  the ratio between the
time-scale of the noise and the coupling between the walkers, we may
explore very different dynamical regimes. This is a relevant degree of
freedom, which permits the control over the transition between different
dynamical evolutions and may be exploited for reservoir engineering,
where noise is shaped to enhance some specific dynamical features.
%%%%%%%%%%%%%%%%%%%%%%%%%%%%%%%%%%%%%%%%%%%%%%%%%%%%%%%%%%%%%%%%%%%%%%%%%%%% 
\section*{ACKNOWLEDGMENTS}
I. S. and P. B. thank A. Beggi for fruitful discussion and suggestions. 
I. S. and C. B. thank A. Buchleitner for useful discussions.
This work has been supported by UniMoRe through FAR2014, by UniMI 
through the H2020 Transition Grant 15-6-3008000-625, and by
EU through the Collaborative Projects QuProCS (Grant Agreement No. 641277).
\bibliography{2pQWn}
\end{document}